\documentclass[aps,amsfonts]{revtex4}

\usepackage{epsfig}

\usepackage{graphicx}
\usepackage{amsmath}
\usepackage{amsbsy}
\begin{document}
\newcommand{\ee}{\end{equation}}
\newcommand{\bb}{\begin{equation}}
\newcommand{\eqb}{\begin{eqnarray}}
\newcommand{\eqf}{\end{eqnarray}}
\def\x{\mathbf{x}}
\def\p{\mathbf{p}}
\def\ho{{\mbox{\tiny{HO}}}}
\def\sc{\scriptscriptstyle}
\title{Neutrino Oscillations in a Minimal CPT Violation Frame }
\author{Paola Arias}
\email{paola.arias@gmail.com} 
\affiliation{Departamento de
F\'{\i}sica, Universidad de Santiago de Chile, Casilla 307,
Santiago, Chile }
 \author{J. Gamboa}
\email{jgamboa55@gmail.com} 
\affiliation{Departamento de
F\'{\i}sica, Universidad de Santiago de Chile, Casilla 307,
Santiago, Chile }

\begin{abstract}
The oscillation neutrino problem in the extended standard model with
minimal number of parameters is considered. The dispersion relations
with explicit neutrino-antineutrino asymmetry are discussed and an
explanation to the MINIBooNE and solar neutrino controversy is
offered. Bounds for the CPT violation symmetry are also found.
\end{abstract}
\maketitle

\section{Introduction}

The analysis and design of neutrino oscillation experiments have been
a very active research field due possible evidence of CPT violation in neutrino physics. Since the controversial LSND result
\cite{LSND}, experiments such as KARMEN \cite{karmen} and specially
MINIBooNE \cite{mini1,mini2} have been designed  to test the LSND
results, which seem to indicate a signal of CPT violation in
neutrino oscillation factories.

As well known, the LSND experiment explored the transition $\bar \nu_\mu
\rightarrow \bar \nu_e$ and providing  a positive appearance of $\bar \nu_e$ over
the background. This last fact  suggest neutrino oscillations at short baseline, with a best fit in
$\Delta m^2_{\mbox{LSND}} \sim 1$ eV$^2$.

Clearly --if this result is true and the CPT theorem holds-- then this is in
contradiction with solar neutrino data where the transition $\nu_e
\rightarrow \nu_\mu$ is favored with a  best fit of $\Delta
m_{SOL}^2 \sim 5.5 \times 10^{-5}$ eV$^2$ and, therefore, the LSND
results  could require an explanation beyond  of conventional
physics.

The MiniBooNE experiment was specially designed to test the LSND
results and, in a first phase, the channel $\nu_\mu \rightarrow \nu_e$ was tested by  exploring energy ranges between $(400-3000)$
MeV. No significant result over the background was reported to high
energies and LSND allowed region was excluded with a 90$\%$CL.
However, a slightly appearance of $\nu_e$ at low energies was
detected \cite{AguilarArevalo:2008rc}.

Motivated by this results, MiniBooNE repeated their measurements,
extending the low energy range to 200 MeV. The excess of $\nu_e$
events was confirmed indicating a positive sign of oscillation, with
an enhancement in the 300 MeV range. The best fit in the low energy
regime is for $E>200 $ MeV and $\Delta m ^2 = 3.14 $ eV$^2$ and
$\sin^22\theta=1.7\times 10^{-3}$. Although this result is
incompatible with LSND-type oscillations, has prompted several
research activity because put us back where we were: incompatibility
between solar and accelerator experiments. In addition to the above,
presently MiniBooNE is running the experiment in the antineutrino
mode, $\bar \nu_\mu \rightarrow \bar \nu_e$, under same  baseline
conditions and energy ranges. Surprisingly, no excess of $\bar
\nu_e$ events have been reported for low energy (nor high) range
\cite{AguilarArevalo:2009xn}.

The MiniBooNE result open the interesting possibility
\cite{Giunti:2007xv,Giunti:2009zz} of consider a tiny CPT violation
in order to explain, in one hand the absence of oscillation in the
antineutrino mode and, on the other hand, the $\Delta m^2_{MB}\sim$
$\mathrm{eV}^{2}$ which is in contradiction with solar neutrino
reports.

In this paper we would like to explore a minimal CPT violating
model \cite{cole}, based on the extended standard model (ESM)
\cite{Colladay:1998fq, Kostelecky:2003xn, Kostelecky:2003cr}.
Basically in this model we take the neutrino ESM sector and then the
corresponding dispersion relations are obtained. These  dispersion
relations have explicit neutrino-antineutrino asymmetry and this
fact allows, in one hand, to reconcile the MiniBoone and solar data
and on the other hand,  to get bounds for the  CPT violating
parameters for this experiments. Other bounds for CPT violation
symmetry in different neutrino oscillation experiments are also
computed and discussed.

\section{CPT violation Lagrangian and equations of motion}

Our starting point is a minimal subset of the SME Lagrangean
\cite{Kostelecky:2003cr,Kostelecky:2003xn}

\bb
\mathcal L= \bar  \nu_a \left(i \Gamma^\mu_{ab} \partial_\mu
-m_{ab}\right) \nu_b,
\ee
 with  $a,b$ flavor indices and
 \bb
\Gamma^\mu_{ab} = \gamma^\mu \delta_{ab}+ e^\mu_{ab}. \ee $m_{ab}$
is the standard CPT conserving mass matrix and we will assume
two-neutrino mixing.\\

In order to preserve rotational invariance, we chose
$e^\mu_{ab} \to e^0_{ab}$ and, therefore, the equations of motion --written in
momentum space-- becomes

\bb \left[\left(E -\vec \alpha \cdot \vec p \right)
\delta_{ab}-\left(m_{ab}-E e_{ab}\right)\gamma^0 \right]\nu_b =0.
\label{dispersion}
\ee

If we choose the chiral representation for the gamma matrices one find that (\ref{dispersion}) yield to the following dispersion relations

\bb
E_\pm ^{(1)}=-\frac{m_1 e_1}{1-e_1^2} \pm
\frac{1}{\sqrt{1-e_1^2}}
\sqrt{m_1^2+p^2+\frac{m_1^2e_1^2}{1-e_1^2}}, \label{esp1}
\ee

\bb
E_\pm ^{(2)}=-\frac{m_2 e_2}{1-e_2^2} \pm
\frac{1}{\sqrt{1-e_2^2}}
\sqrt{m_2^2+p^2+\frac{m_2^2e_2^2}{1-e_2^2}},  \label{esp2}
\ee
where the $\pm$ signs denote particle ($+$) and antiparticle ($-$) and
$m_{1,2}$ and $e_{1,2}$ correspond to the eigenvalue of matrix $m$
and $e$, respectively. Note that (\ref{esp1}) and (\ref{esp2}) have
particle-antiparticle asymmetry and, therefore, explicit CPT
violation.

The mixing angle is fixed to Large Mixing approximation (LMA)
because the elements off diagonal in $m$ and $e$ matrices were set
to zero. However, if we consider a more general set of matrices $m$
and $e$ with the same basis of eigenvectors, then the mixing angle
would be not fixed and has an energy dependence, with different
values for neutrino and antineutrinos.\\

Next step is to expand the above dispersion relations in powers of $m^2/p^2$, thus

\bb
E_a^\pm=\mp
m_ae_a+p\left(1+\frac{m_a^2}{2p^2}+\frac{e_a^2}2\right),
\label{ener}
\ee
where $p(1+m^2/p^2)$ correspond to the first terms of the expansion $\sqrt{p^2+m^2}$ and the following ones are Lorentz and CPT violating contributions (see
{\it e.g.} \cite{lisi,bahcall}).  We finalize this section by noticing that the CPT violation have been considered only at the kinetic level and this fact is enough  to generate  (\ref{ener}).

\section{Oscillation probability in CPT violation  frame}

Recall that the probability of oscillation between two different
species of neutrinos in a CPT conserving scenario, say
$\nu_{\alpha}$ and $\nu_{\beta}$ is given by

\bb 
P_{\nu_\alpha\rightarrow \nu_\beta}= \sin^2 2\theta ~\sin^2
\left(\frac{L}{L_0}\right), \ee 
where $L$ is the length from the
production source to the detector and $L_0$ is the so-called
oscillation length, given by

\bb L_0 = \frac{4\pi E}{\Delta m^2_{ij}}.\ee In order to observe
oscillations, it must be satisfied that $L/L_0 \geq 1$.

Maximal sensitivity to oscillations, and therefore $\Delta
m^2_{ij}$, is obtained when the setup of the experiment is such that
\bb \frac{E}L \approx \Delta m_{ij}^2.\ee

In our CPT violation scenario, the oscillation probability between two
different neutrino species is computed to be \bb
P_{\nu_\alpha\rightarrow \nu_\beta}= \sin^2 2\theta \sin^2 \left(
\Delta E_{ab}L \right), \ee where $\Delta E_{ab}= E_a-E_b$. Thus,
using  the dispersion relation, eq. (\ref{ener}) we obtain

\bb 
P_{\nu_\alpha\rightarrow \nu_\beta}= \sin^2 2\theta ~\sin^2
\left(\frac{\Delta m_{12}^2}{2E}L+\frac{\Delta
e_{12}^2E}2L+(m_2e_2-m_1e_1)L\right), 
\ee 
and for antineutrinos

\bb 
P_{\bar\nu_\alpha\rightarrow \bar \nu_\beta}= \sin^2 2\bar
\theta ~\sin^2 \left(\frac{\Delta m_{12}^2}{2E}L+\frac{\Delta
e_{12}^2E}2L-(m_2e_2-m_1e_1)L\right). 
\ee

The bar over the mixing angle accounts for the possibility of
consider a different mixing angle for particle and antiparticle, as
was discussed before.\\ \noindent  The first term in sine argument
corresponds to the standard difference of squared masses over
energy. The rest corresponds to the
contribution of CPT violation.

The term independent of the energy, of the form $\eta^0_{ab} =m_ae_a
- m_be_b$, has been constrained in \cite{lisi} using bounds for
atmospheric neutrinos to $ \eta^0_{23} < 10^{-23} $ GeV . Meanwhile,
from solar neutrino data \cite{bahcall} the corresponding bound is
$\eta^0_{12} < 10^{-21} $ GeV. As expected, this term is well
attenuated, and has almost none influence in the oscillation
phenomena.\\

The oscillation length is modified, adding a term which becomes
significant at high  energies

\bb L_0^{CPTV }= \frac{1}{\frac{\Delta m^2}{4\pi E}+\frac{\Delta e^2
E}2 }. \ee

Neutrino transitions can be observed if 
$L\geq L_0^{CPTV}$ and, therefore, one has 

\bb 
\Delta m^2_{ij} X \left(1+\frac{\Delta e_{ij}^2 E^2}{\Delta
m^2_{ij}}\right)\geq 1, \label{param}
\ee 
where $X=L/E$. Experimental
setups are arranged such that $\Delta m^2_{ij} X \sim 1$. Using this
expression
we can find bounds for $\Delta e^2_{ij}$ from different experimental data.\\

\subsection{Bounds on $\Delta e_{ij}^2$}

According to the MiniBooNE setup, the distance from the source to the
detector is $L=541 $m and the range of energy for observed
oscillations is 300-475 MeV, with $\Delta m_{12}^2$ of the order of
1 $eV^2$. Using formulae (\ref{param}) we can find a bound of
$\Delta e_{12}^2$

\bb 
\Delta e^2_{12} \leq  10^{-18}, \ \ \ \ \ {\text{for \ \ }
\Delta m_{12}^2 \sim 1 \ \text{eV}^2 }.
\ee

Using this bound for $\Delta e^2_{12}$,  is clear that for solar
neutrino oscillations the CPT violation is well suppressed. Indeed,  the solar neutrino data are $E=0.8 $ MeV and $\Delta
m_{12}^2\sim 10^{-5}$ and, therefore the coefficient must satisfy 
\bb 
\frac{\Delta
e_{12}^2 E^2}{\Delta m^2_{12}}<<1 .
\ee

The absence of neutrino oscillations in MiniBooNE for the range of energies
over 500 MeV is very natural; the coefficient $X <<1$ and besides,
$\frac{\Delta e_{ij}^2 E^2}{\Delta m^2_{ij}} \leq 1$ in the range of
energy between
500-3000 MeV. Thus, the oscillation is suppressed.\\

If we try to conciliate the solar and MiniBooNE masses, one should set
$\Delta m^2_{12}= \Delta m^2_{SOL} =\Delta m^2_{MB} \sim 10^{-5}$
eV$^2$, in which case the bound for $\Delta e^2_{12}$, using the
above formula, becomes 

\[ 
|\Delta e_{12}^2 |\leq 10^{-17} .
\]

Using atmospheric neutrino oscillation data we can constraint
the parameter $\Delta e_{23}^2$. Indeed, using existing data for $\Delta m^2_{23} \sim 10^{-3} eV^2$, $X\sim 10^{-3} eV^2$, and the neutrino
energy in multi-GeV events  $E\sim 3$ GeV, we obtain the bound 

\bb
\Delta e^2_{23} \leq 10^{-21}.
\ee

\section{Conclusions}

In this paper we have proposed a minimal version of the ESM containing two free parameters responsible of the CPT violation in the neutrino sector. In connection with the oscillation probability, we found that the oscillation length is modified by the inclusion of the CPT violation parameters (see also \cite{Fogli:1999fs}) with a minimal number of free parameters. Thus, the simplified model studied here allowed us to
constraint the parameters that determines the violation of symmetry.

From  the MiniBooNE results for low-energy neutrino oscillation we find the bound $\Delta e^2_{12}=e^2_1-e^2_2 \leq  10^{-18}$, however an agreement between MiniBooNE and solar neutrino data can be reached if  $|\Delta e^2_{12}| $ is one order of magnitude below {\it i.e.}  $ \Delta e^2_{12} \leq  10^{-17}$.  For completeness, the bound for $\Delta e^2_{23}$ was also obtained, using the atmospheric neutrino oscillation data. The bound is $\Delta e^2_{23} \leq 10^{-21}$.

It is interesting to note that from the dispersion relation obtained
in the paper a possible scenario of neutrino-antineutrino
oscillation also could be considered. For example, the difference
$|E_{a}^+-E_a^-|\approx 2m_a e_a$ tell us that, if the mixing angle is
not suppressed, then an oscillation could appear as a CPT violation effect \cite{Hollenberg:2009tr} (the
same conclusion would be reached by considering an oscillation between
neutrino-antineutrino of different species). 

\vspace{0.3 cm}

\noindent\underline{Acknowledgements}:  We would like to thank J. L\'opez-Sarri\'on by fruitful discussions. This work was partially
supported by FONDECYT-Chile grant-1095106 and DICYT (USACH).



\end{document}